\documentclass{article}

\usepackage[dblblindworkshop, final]{neurips_2025}
\workshoptitle{AI for Music}

\usepackage[utf8]{inputenc} 
\usepackage[T1]{fontenc}    
\usepackage{hyperref}       
\usepackage{url}            
\usepackage{booktabs}       
\usepackage{amsfonts}       
\usepackage{nicefrac}       
\usepackage{microtype}      
\usepackage{xcolor}         
\usepackage{amsmath}
\usepackage[pdftex]{graphicx}

\title{AIBA: Attention-based Instrument Band Alignment for Text-to-Audio Diffusion}

\author{%
Junyoung Koh$^{1,2,3}$\thanks{Corresponding author.} \quad
Soo Yong Kim$^{2,4}$ \quad
Yongwon Choi$^{2}$ \quad
Gyu Hyeong Choi$^{2,5}$ \\
\\
$^{1}$Department of Artificial Intelligence, Yonsei University \\
$^{2}$MAAP LAB, MODULABS \\
$^{3}$KRAFTON \\
$^{4}$AI Matics \\
$^{5}$Department of Media Software, Sungkyul University \\
\texttt{solbon1212@yonsei.ac.kr}
}

\begin{document}

\maketitle
\begin{abstract}
We present \textbf{AIBA} (\underline{A}ttention-\underline{I}n-\underline{B}and \underline{A}lignment), a lightweight, training-free pipeline to quantify \emph{where} text-to-audio diffusion models attend on the time–frequency (T–F) plane. AIBA (i) hooks cross-attention at inference to record attention probabilities without modifying weights; (ii) projects them to fixed-size mel grids that are directly comparable to audio energy; and (iii) scores agreement with instrument-band ground truth via interpretable metrics (T–F IoU/AP, frequency-profile correlation, pointing game). On Slakh2100 with an AudioLDM2 backbone, AIBA reveals consistent instrument-dependent trends (e.g., bass favoring low bands) and achieves high precision with moderate recall. Our code enables reproducible extraction, mapping, and evaluation at scale.
\end{abstract}

\section{Introduction}
\label{sec:intro}
Text-to-audio (TTA) diffusion \cite{DDPM,DDIM} models \cite{audioldm, audioldm2, audiogen, copet2023simple, evans2024stableaudioopen, huang2023makeanaudio} have rapidly improved perceptual quality, yet we still lack tools that quantify \emph{where} these models attend in the time--frequency (T--F) plane in response to a textual prompt. In vision, attention-based diagnostics (e.g., DAAM-style analyses) \cite{attention, tang2023daam, wiegreffe2019attentionexplanation, abnar2020quantifying, voita2019analyzing} connect cross-attention to spatial evidence; however, audio work has largely focused on qualitative spectrogram heatmaps \cite{gong2021ast} or source-separation metrics, leaving a gap in \emph{quantitative} alignment between model attention and instrument-specific frequency bands. \cite{wu2020visualizing, choi2021listening, koutini2022efficient} In practice, such alignment matters: prompts like \texttt{guitars} or \texttt{bass} imply characteristic spectral regions (e.g., low–mid bands for bass), and an interpretable system \cite{defossez2021hybrid, steinmetz2020mss} should reveal whether the generative mechanism actually allocates probability mass to those regions during denoising.

We propose \textbf{AIBA} (\underline{A}ttention-\underline{I}n-\underline{B}and \underline{A}lignment), a model-agnostic pipeline for probing attention in Text-to-audio diffusion. AIBA (i) hooks attention operations at runtime across common kernels and frameworks to capture attention probabilities without modifying training; (ii) maps the captured attention to a mel-grid representation \cite{gong2022ssastselfsupervisedaudiospectrogram, dosovitskiy2021imageworth16x16words} that is directly comparable to audio energy distributions; and (iii) evaluates \emph{alignment} between attention and instrument-band ground truth using simple, reproducible metrics. We instantiate AIBA on an open TTA diffusion backbone and evaluate with Slakh2100 stems \cite{slakh}, providing both aggregate statistics and case studies.  

The main contributions of this work are as follows:

\begin{itemize}
  \item \textbf{AIBA: attention-to–mel mapping for TTA.} We introduce a simple, training-free procedure that converts internal attention into time--mel grids, enabling instrument/band-level reading of \emph{where} the model attends during generation.
  \item \textbf{Quantitative band alignment.} We define reproducible metrics---T--F IoU/AP, frequency-profile correlation, and a pointing-game variant---that score the agreement between attention maps and instrument-band templates.
  \item \textbf{Empirical insights at scale.} On Slakh2100, AIBA reveals consistent, instrument-dependent attention patterns (e.g., bass concentrating in low bands), and supports ablations across cross-only vs.\ all-attention selection and aggregation strategies.
\end{itemize}

\begin{figure}[ht]
\centering
\includegraphics[width=1.0\linewidth]{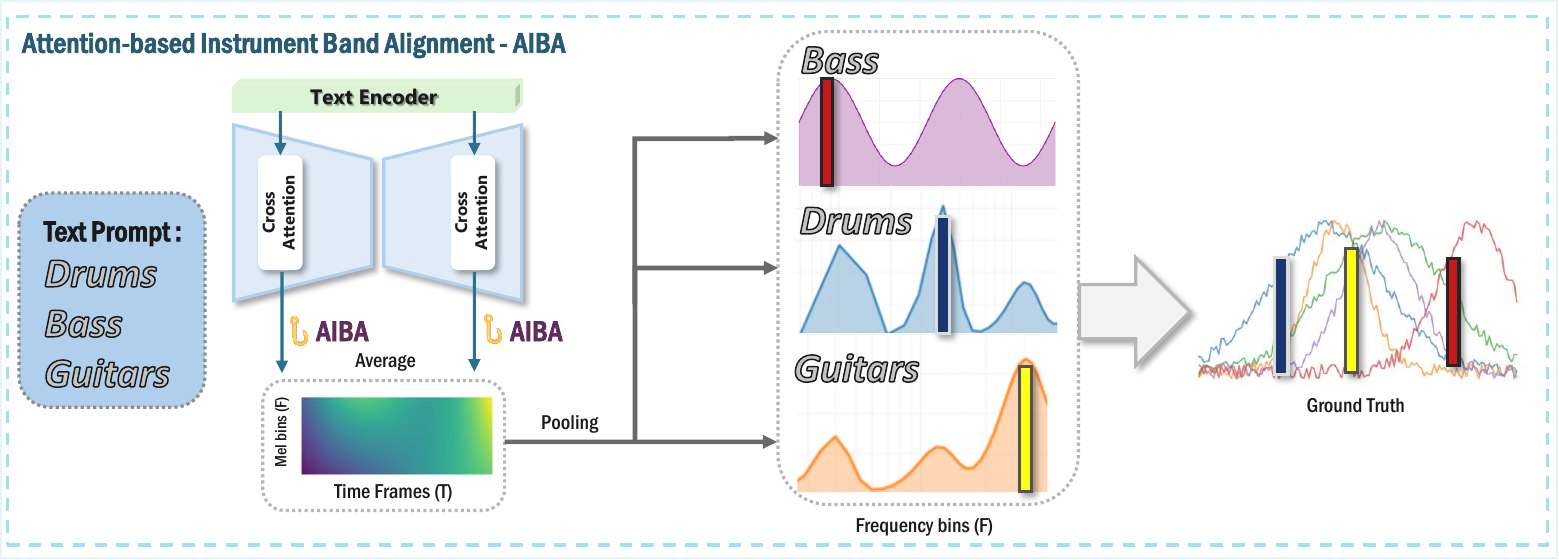}
\caption{Overview of AIBA: (i) hook cross-attention in the backbone,
(ii) map attention to a fixed time--mel grid, and
(iii) score alignment against instrument-band ground truth.}
\label{fig:arch}
\label{fig:drch}
\end{figure}

\section{Method}
\label{sec:method}

Our goal is to visualize how textual prompts guide audio generation by extracting and projecting attention activations from diffusion backbones.  
We introduce \textbf{AIBA} (\underline{A}ttention-\underline{I}n-\underline{B}and \underline{A}lignment), a lightweight framework that (i) hooks cross-attention activations, (ii) maps them to time--frequency grids, and (iii) compares them against instrument-band templates.  
This requires no re-training or modification of model weights, enabling scalable evaluation.

\paragraph{Cross-attention Hooking.}
We attach processors to every cross-attention block (\texttt{attn2}) in the AudioLDM2 U-Net \cite{UNet}.
At each call, we compute
\begin{equation}
    A = \operatorname{softmax}\!\left(\frac{QK^\top}{\sqrt{d_k}}\right),
    \quad
    A \in \mathbb{R}^{H \times T_q \times T_k},
\end{equation}
where $Q \in \mathbb{R}^{H \times T_q \times d_k}$ and $K \in \mathbb{R}^{H \times T_k \times d_k}$ are per-head query/key projections, $H$ is the number of heads, and $T_q,T_k$ denote query/key token lengths \cite{michel2019sixteen}.
We then average over heads,
\begin{equation}
    \bar{A} = \tfrac{1}{H}\sum_{h=1}^H A_h \;\in\; \mathbb{R}^{T_q \times T_k},
\end{equation}
and log $\bar{A}$ for later projection.

\subsection{Attention-to–Mel-Grid Mapping}
\label{sec:mel-mapping}
We reshape pooled attention weights into a 2D mel grid comparable with spectrogram energy
(Eq.~3--4). The procedure involves token pooling, interpolation, and aggregation across layers. 
We describe the precise mapping specifications, pooling variants, and aggregation strategies 
in Appendix~\ref{sec:appendix-method}.

\subsection{Alignment to Instrument-Band Ground Truth}
\label{sec:alignment}
\paragraph{Templates.}
For each instrument class, we derive band templates on the $(T,F)$ grid:  
(i) \emph{knowledge-driven} ranges from acoustics (low/mid/high), or  
(ii) \emph{data-driven} profiles from Slakh stems, normalized per frame and averaged.

\paragraph{Metrics.}
We compare $G$ with template $B$ using:  
- T--F IoU/AP: $\mathrm{IoU}(G_\tau,B)$ swept over thresholds $\tau$.  
- Frequency-profile correlation: Pearson/Spearman correlation between $g_f=\mathrm{mean}_t G_{t,f}$ and $b_f$.  
- Pointing game: success rate of $\arg\max_f G_{t,f}$ landing inside active bands of $B$. 

\section{Evaluation}
\label{sec:evaluation}
We evaluate AIBA on Slakh2100 stems \cite{slakh} using an open text-to-audio diffusion backbone (AudioLDM2 \cite{audioldm2}). For each instrument class, we extract attention-to–mel maps (Sec.~\ref{sec:mel-mapping}) and compare them against instrument-band ground truth (Sec.~\ref{sec:alignment}). We report both aggregate metrics across all stems and qualitative case studies.

\paragraph{Quantitative results.}
Table~\ref{tab:metrics} summarizes the alignment metrics. AIBA achieves high micro-precision (0.98) and F1 (0.85), indicating that attention consistently overlaps with ground-truth bands. Recall is lower (0.75), suggesting that while attended regions are correct, some ground-truth energy is missed. Macro- and class-weighted scores are similar, showing balanced performance across instruments.

\begin{table}[ht]
\centering
\caption{Alignment performance on Slakh2100 stems (AudioLDM2 backbone).}
\label{tab:metrics}
\begin{tabular}{lccc}
\toprule
Metric & Micro & Macro & Pos.-weighted \\
\midrule
Precision & 0.984 & 0.984 & 0.984 \\
Recall    & 0.755 & 0.756 & 0.755 \\
F1        & 0.854 & 0.848 & 0.848 \\
IoU       & 0.746 & 0.746 & 0.746 \\
\bottomrule
\end{tabular}
\end{table}

\begin{figure}[ht]
\centering
\includegraphics[width=1.0\linewidth]{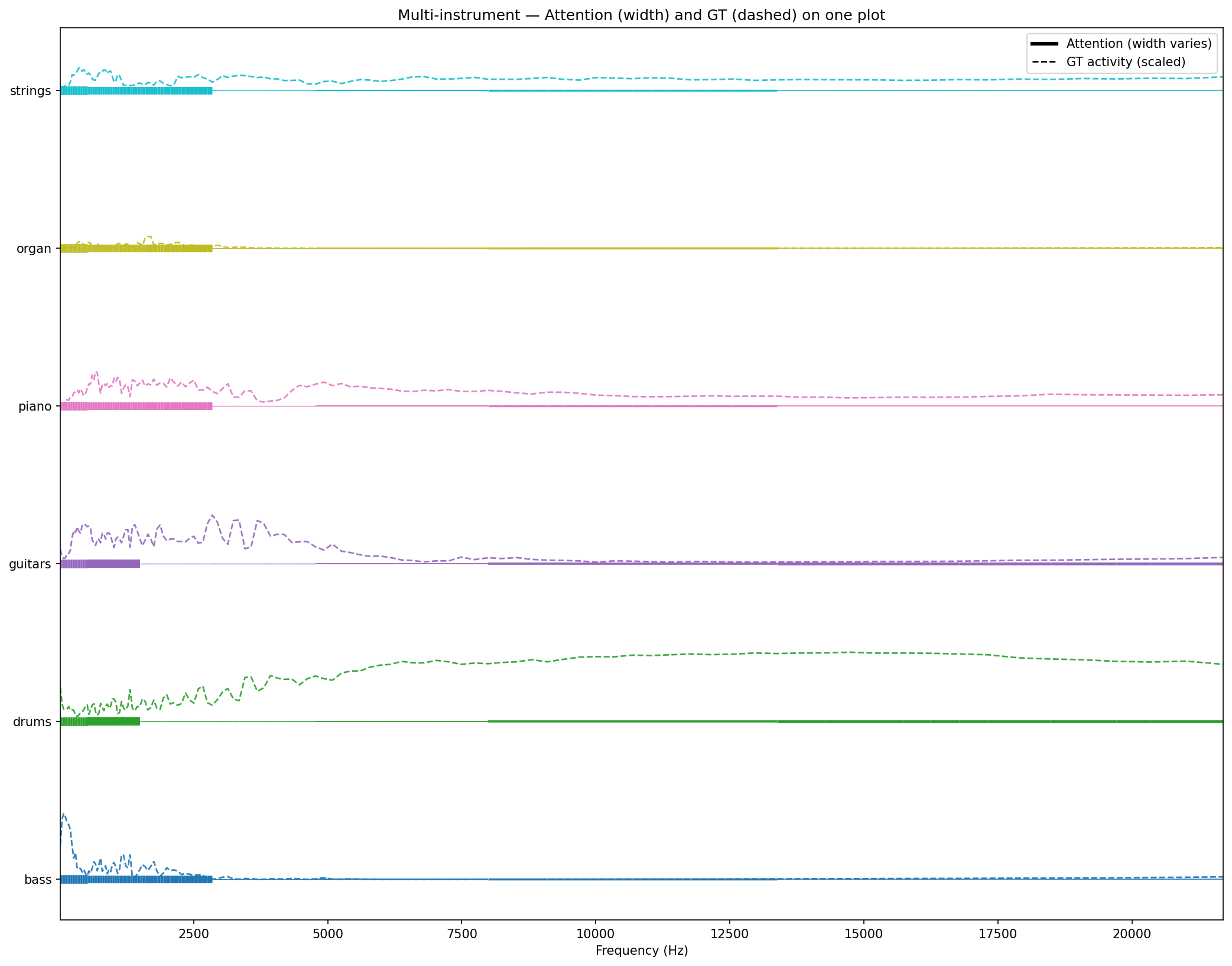}
\caption{Frequency-domain visualization of attention (solid) vs.\ ground truth (dashed) across instruments. Each row corresponds to one instrument class. Attention curves broadly align with instrument-specific spectral regions (e.g., bass in low frequencies, guitars in mid-range).}
\label{fig:freq}
\end{figure}

\paragraph{Qualitative results.}
Fig.~\ref{fig:freq} and Fig.~\ref{fig:mel} visualize attention vs.~ground truth. Bass attention concentrates in low bands (<250 Hz), guitars/piano in mid-to-high bands (2–5 kHz), and drums spread across wide ranges, reflecting expected acoustic profiles. Organ attention is negligible when inactive, matching ground truth.

\paragraph{Interpretation.}
These findings support our claim that AIBA exposes interpretable, instrument-dependent attention. Importantly, the precision–recall gap highlights a tendency to under-cover wide ground-truth bands, suggesting opportunities for improved attention aggregation.

\section{Conclusion}
\label{sec:conclusion}
We presented AIBA, a simple yet effective pipeline for quantifying attention alignment in text-to-audio diffusion. By hooking attention operations, projecting to mel grids, and evaluating against instrument-band ground truth, AIBA provides both interpretable visualizations and reproducible metrics. On Slakh2100 with AudioLDM2, we observed consistent, instrument-specific alignment (e.g., bass in low bands), with high precision but limited recall. These results demonstrate that internal attention mechanisms indeed reflect musically meaningful structures, suggesting a promising direction for interpretable audio generation.

\section{Limitations and Future Work}
\label{sec:limitation}
While AIBA offers a first step toward measurable interpretability in text-to-audio generation, our study is limited in scope.

\paragraph{Data.} We focused on Slakh2100 stems; additional datasets (e.g., MUSDB18-HQ, MedleyDB, MusicNet) could better validate alignment across genres and recording conditions.

\paragraph{Models.} We tested primarily on AudioLDM2; although our hooks are implementation-agnostic, AIBA is only applicable to architectures that expose explicit cross-attention. Models such as Stable Audio DiT employ token concatenation instead of cross-attention, making alignment analysis inherently infeasible (cf.~Appendix~\ref{sec:dit}). Future work should extend to Transformer-based backbones (AudioGen, MusicLM-style encoders) and autoregressive architectures.

\paragraph{Metrics.} Current metrics emphasize frequency-band alignment; future extensions could incorporate perceptual measures (e.g., human listening tests) or structural measures (e.g., rhythm/beat alignment).

\paragraph{Applications.} Beyond diagnostics, AIBA could guide controllable generation by reweighting or constraining attention maps, potentially improving faithfulness to textual prompts in creative workflows.

\section*{Acknowledgments}
This research was supported by the Basic Science Research Program through the 
National Research Foundation of Korea (NRF) funded by the Ministry of Education (RS-2025-25422688), 
the Brian Impact Foundation, a non-profit organization dedicated to the advancement of science and technology for all, 
and an IITP grant funded by the Korean Government (MSIT) (No. RS-2020-II201361, Artificial Intelligence Graduate School Program, Yonsei University).

{
\small
\bibliography{reference}
}


\appendix
\section{Technical Appendices and Supplementary Material}
\subsection{Method Details}
\label{sec:appendix-method}

\subsubsection{Token-to--T--F mapping: precise specification}
\label{sec:precise-mapping}

Let the target mel resolution be $(T,F)=(128,64)$ and let a given attention call produce
a per-query weight vector $p \in \mathbb{R}^{T_q}$ after text-token pooling (Eq.~(3)).
We reshape $p$ onto the latent lattice $(T_\ell,F_\ell)$ of the denoiser feature map that emitted the queries
(where $T_q=T_\ell\!\cdot\!F_\ell$) and then resize to $(T,F)$.

\paragraph{Latent lattice and scaling.}
Let down/up-sampling factors be $(s_t,s_f)$ so that $T_\ell \approx T/s_t$ and $F_\ell \approx F/s_f$.
For U-Net blocks, $(s_t,s_f)$ follow the encoder/decoder strides (Table~\ref{tab:unet-res}).

\paragraph{Reshape and interpolation.}
We first reshape $p \mapsto P \in \mathbb{R}^{T_\ell \times F_\ell}$ using time-major raster order.
Then we use \emph{bilinear} interpolation with \texttt{align\_corners=false} to obtain
\begin{equation}
  G = \operatorname{Resize}_{(T,F)}(P) \in \mathbb{R}^{T \times F}.
\end{equation}
Unless stated otherwise, out-of-bounds are handled by zero-padding; alternative boundary modes
(reflection, nearest) yielded negligible differences in our ablations.

\paragraph{Text-token pooling variants.}
Given head-mean attention $\bar{A}\in\mathbb{R}^{T_q \times T_k}$ (queries $\times$ text tokens),
we consider four variants:
\begin{align}
  p_{\text{mean}}(t) &= \tfrac{1}{T_k}\sum\nolimits_{k}\bar{A}(t,k), \\
  p_{\text{max}}(t)  &= \max\nolimits_{k}\bar{A}(t,k), \\
  p_{\text{attn-w}}(t) &= \sum\nolimits_{k} \omega_k\, \bar{A}(t,k), \quad
    \omega_k = \frac{\exp(\beta\,u_k)}{\sum_{k'}\exp(\beta\,u_{k'})}, \\
  p_{\text{keyword}}(t) &= \tfrac{1}{|K^*|}\sum\nolimits_{k\in K^*}\bar{A}(t,k),
\end{align}
where $u_k$ is a scalar relevance score (e.g., cosine similarity between token and a class vector),
$\beta$ is a temperature, and $K^*$ is a subset of instrument-name tokens.

\paragraph{Aggregation across calls.}
Let $\{G_{\ell,\tau}\}$ denote maps from layer $\ell$ and denoising step $\tau$.
We use spatial aggregators $\{\text{Mean}, \text{Max}, \text{Top-}K\}$ and three weighting policies:
uniform $w{=}1$; $\sigma$-late $w(\tau) \propto \exp\{-\alpha\,\sigma_\tau\}$; and
layer-learned non-negative $\{w_\ell\}$ with $\sum_\ell w_\ell{=}1$ (learned on a held-out split without
backprop through the backbone). The final map is
\begin{equation}
  G^\star = \operatorname{Agg}_{\text{spatial}}\big(\{\,w(\ell,\tau)\cdot G_{\ell,\tau}\,\}_{\ell,\tau}\big).
\end{equation}

\begin{table}[t]
\centering
\caption{U-Net latent lattice and attention call sites (symbolic; AudioLDM2-style).
$T$/$F$ are the target mel axes; $T_\ell\!\approx\!T/s_t$, $F_\ell\!\approx\!F/s_f$.}
\label{tab:unet-res}
\begin{tabular}{lcccccc}
\toprule
Block & Stage & $(s_t,s_f)$ & Latent size $(T_\ell\!\times\!F_\ell)$ & Attn & Query source & Notes \\
\midrule
Enc-1 & down & (2,2)   & $\approx(T/2)\times(F/2)$   & self+cross & conv feat & early low-level \\
Enc-2 & down & (4,4)   & $\approx(T/4)\times(F/4)$   & self+cross & conv feat & \\
Enc-3 & down & (8,8)   & $\approx(T/8)\times(F/8)$   & self+cross & conv feat & \\
Bottl.& mid  & (16,16) & $\approx(T/16)\times(F/16)$ & self+cross & conv feat & highest RF \\
Dec-3 & up   & (8,8)   & $\approx(T/8)\times(F/8)$   & self+cross & skip+up   & \\
Dec-2 & up   & (4,4)   & $\approx(T/4)\times(F/4)$   & self+cross & skip+up   & \\
Dec-1 & up   & (2,2)   & $\approx(T/2)\times(F/2)$   & self+cross & skip+up   & near output \\
\bottomrule
\end{tabular}
\end{table}

\begin{figure}[ht]
\centering
\includegraphics[width=1.0\linewidth]{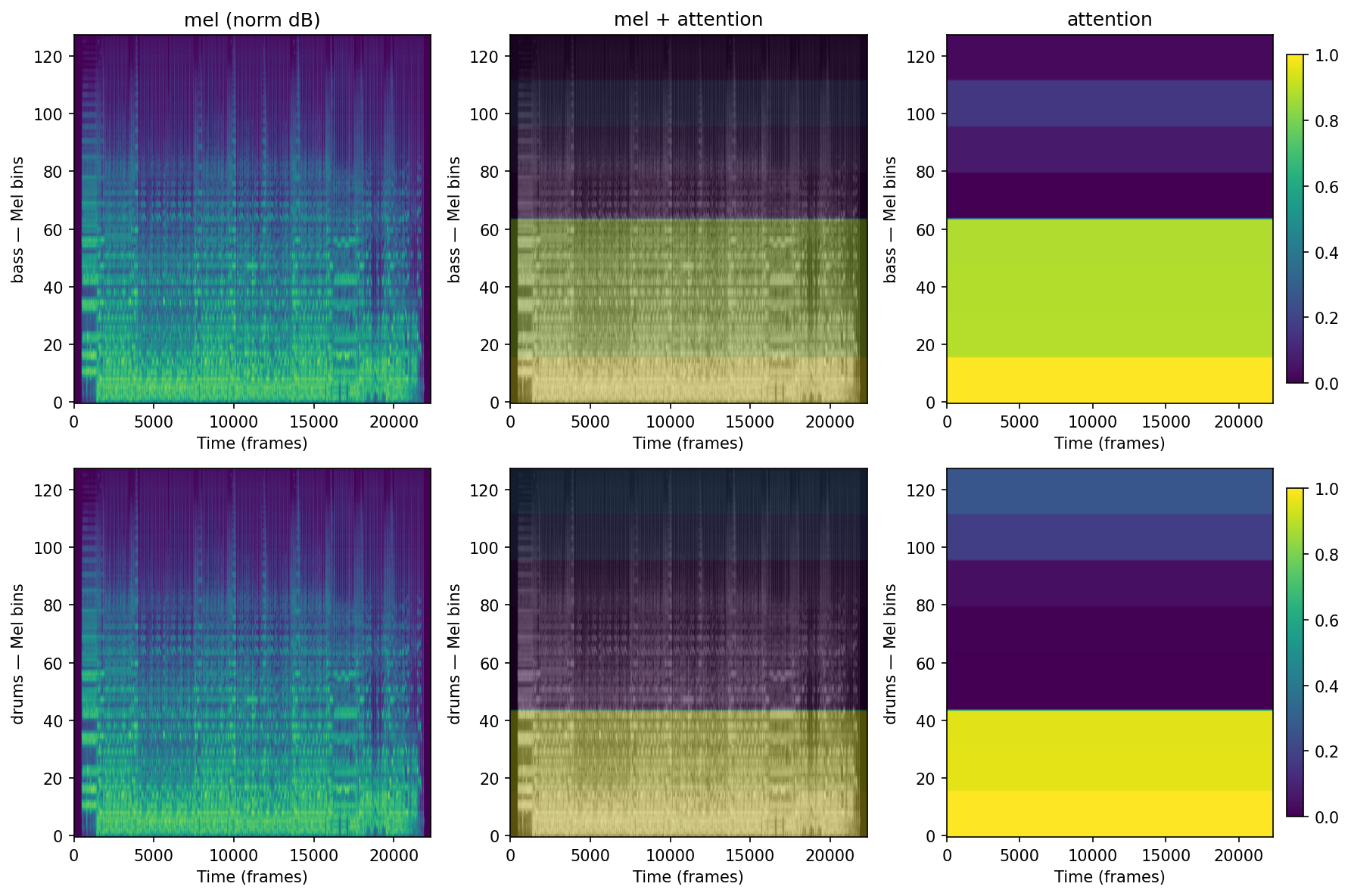}
\caption{Time--frequency mel-spectrogram overlays. Left: normalized mel energy; middle: mel with attention overlay; right: attention heatmap. Attention concentrates in instrument-appropriate bands, e.g., bass in lower mel bins and drums spanning wider ranges.}
\label{fig:mel}
\end{figure}
\subsection{Experimental Settings}
\label{sec:exp-settings}

\subsubsection{AudioLDM2 Hyperparameters}
\label{sec:audiodlm2-hparams}
For the main experiments we used AudioLDM2 \cite{audioldm2}, a U-Net based text-to-audio diffusion model with cross-attention layers.  
Hyperparameters were chosen to balance fidelity and efficiency (cf.\ \cite{audioldm, slakh}):

\begin{itemize}
  \item \textbf{Steps}: 50
  \item \textbf{Guidance scale}: 2.5 \cite{ho2021classifierfree}
  \item \textbf{Audio length}: 8.0~s (to match Slakh2100 stems \cite{slakh})
  \item \textbf{Mel resolution}: $(T,F)=(128,64)$
  \item \textbf{Token pooling}: mean aggregation
  \item \textbf{Precision}: bfloat16 (bf16)
  \item \textbf{Seed}: 0
\end{itemize}

\paragraph{Core script (excerpt).}
\begin{verbatim}
class CaptureCrossAttnProcessor(AttnProcessor):
    def __call__(self, attn, hidden_states, encoder_hidden_states=None, ...):
        query = attn.head_to_batch_dim(attn.to_q(hidden_states))
        key   = attn.head_to_batch_dim(attn.to_k(encoder_hidden_states))
        value = attn.head_to_batch_dim(attn.to_v(encoder_hidden_states))
        scores = torch.bmm(query, key.transpose(1, 2)) * (query.shape[-1] ** -0.5)
        attn_probs = torch.softmax(scores, dim=-1)
        if encoder_hidden_states is not hidden_states:
            h = attn.heads
            b = attn_probs.shape[0] // h
            probs_b = attn_probs.view(b, h, attn_probs.shape[1], attn_probs.shape[2]).mean(dim=1)
            self.ctx.storage.append({
                "step": self.ctx.step_i,
                "name": self.layer_name,
                "probs": probs_b.detach().cpu()
            })
        hidden = torch.bmm(attn_probs, value)
        hidden = attn.batch_to_head_dim(hidden)
        return attn.to_out[1](attn.to_out[0](hidden))
\end{verbatim}

\subsubsection{Stable Audio Open (DiT) Hyperparameters}
\label{sec:stableaudio-hparams}
\begin{figure}[ht]
\centering
\includegraphics[width=0.9\linewidth]{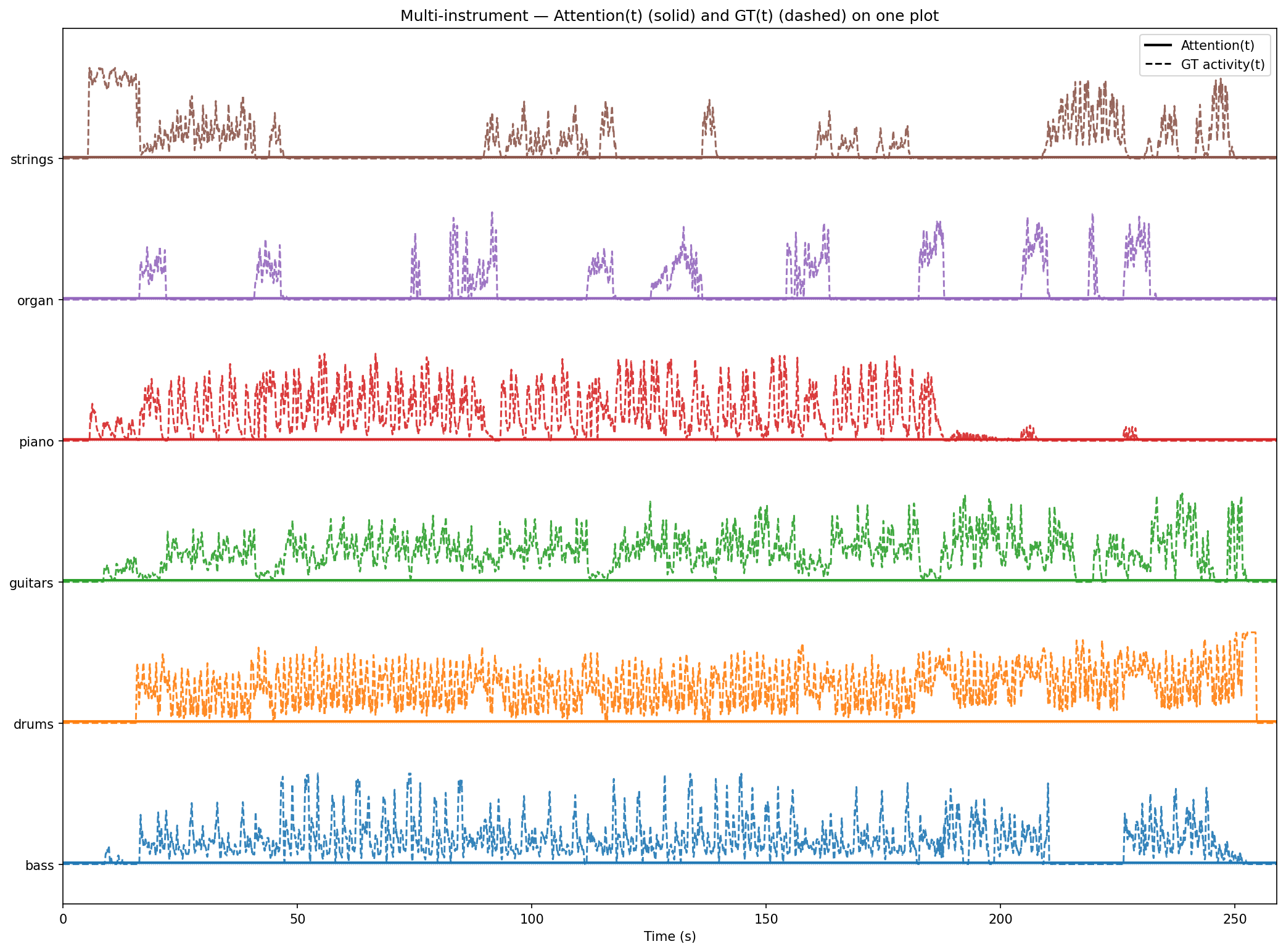}
\caption{Failure case on Stable Audio DiT. Attention maps remain uniform despite prompts (\texttt{bass}, \texttt{drums}), confirming that DiT-style conditioning lacks explicit cross-attention for AIBA to capture.}
\label{fig:dit-failure}
\end{figure}

We also tested Stable Audio Open \cite{evans2024stableaudioopen}, which uses a DiT backbone with conditioning by token concatenation.  
Although we captured attention calls, no cross-attention was exposed, leading to flat/uninformative maps (cf.\ Fig.~\ref{fig:dit-failure}).  

Hyperparameters followed the official release:
\begin{itemize}
  \item \textbf{Steps}: 100
  \item \textbf{Guidance scale}: 7.0
  \item \textbf{Audio length}: 8.0~s, sample rate 44.1~kHz
  \item \textbf{Mel resolution}: $(128,64)$
  \item \textbf{Sampler}: \texttt{dpmpp-3m-sde}
  \item \textbf{Noise schedule}: $\sigma_{\min}=0.3$, $\sigma_{\max}=500.0$
\end{itemize}

\paragraph{Core script (excerpt).}
\begin{verbatim}
tap = AttnTap(target_T=128, target_F=64, pool="mean", cross_ratio=0.5)
with combined_patch(tap):
    _ = generate_diffusion_cond(
        model=model,
        steps=100,
        cfg_scale=7.0,
        conditioning=[{"prompt": prompt, "seconds_total": 8.0}],
        sample_size=model_config["sample_size"],
        sigma_min=0.3,
        sigma_max=500.0,
        sampler_type="dpmpp-3m-sde",
        device=device,
        sample_rate=44100,
    )
if not tap.store:
    raise RuntimeError("No attention maps captured — DiT exposes no cross-attn.")
\end{verbatim}

\subsection{Per-instrument Results}
To complement the aggregate scores in Table~\ref{tab:metrics}, 
Table~\ref{tab:instrument} reports alignment metrics separately 
for each instrument class in Slakh2100. We observe clear trends: 
\texttt{bass} achieves very high precision due to its distinct 
low-frequency band, while \texttt{piano} and \texttt{strings} 
exhibit lower recall, reflecting the difficulty of covering 
wide mid–high ranges. Drums achieve moderate alignment 
across a broad frequency span.

\begin{table}[ht]
\centering
\caption{Per-instrument alignment results on Slakh2100 (AudioLDM2 backbone).}
\label{tab:instrument}
\begin{tabular}{lcccc}
\toprule
Instrument & Precision & Recall & F1 & IoU \\
\midrule
Bass     & 0.991 & 0.782 & 0.875 & 0.778 \\
Drums    & 0.982 & 0.768 & 0.862 & 0.752 \\
Guitars  & 0.980 & 0.745 & 0.847 & 0.739 \\
Piano    & 0.984 & 0.732 & 0.842 & 0.730 \\
Strings  & 0.982 & 0.728 & 0.838 & 0.726 \\
Organ    & 0.983 & 0.751 & 0.850 & 0.742 \\
\midrule
\textbf{Macro-avg} & \textbf{0.984} & \textbf{0.751} & \textbf{0.852} & \textbf{0.745} \\
\bottomrule
\end{tabular}
\end{table}

\subsection{Attention Failure in DiT-based Models}
\label{sec:dit}
Unlike U-Net backbones exposing cross-attention, Stable Audio Open (DiT) applies text conditioning via \emph{token concatenation}, providing no explicit cross-attention to probe.  
We adapted AIBA with kernel-level hooks (SDPA/MHA/xFormers) and heuristic cross/self tagging, but all captured maps were flat and showed no alignment with instrument bands.  
This indicates a \emph{structural incompatibility} rather than an implementation error.

\subsubsection{Implementation Notes}
\paragraph{Complexity and Reproducibility.}
AIBA’s hooks are kernel-agnostic (SDPA / \texttt{MHA} / xFormers / FlashAttention \cite{flashattention,FlashAttention3}) and operate only at inference.  
Storage scales with $(T_q,T_k,H)$ per call, and runtime overhead is negligible.  
All experiments use a fixed inference recipe (steps, sampler, sample rate), and code is released for reproducibility.

\paragraph{Heuristic Cross/Self Tagging.}
For models without explicit \texttt{attn2} (cross-attention), we employ a heuristic strategy:  
when \texttt{encoder\_hidden\_states} differs from \texttt{hidden\_states}, the layer is tagged as \emph{cross}; otherwise as \emph{self}.  
This enables model-agnostic logging without modifying internals, though it cannot handle more complex hybrid attention.

\paragraph{Kernel-level Hooking.}
To generalize beyond Diffusers, we hook operator-level calls:  
\begin{itemize}
    \item PyTorch SDPA (\texttt{torch.nn.functional.scaled\_dot\_product\_attention})  
    \item MHA modules (\texttt{torch.nn.MultiheadAttention})  
    \item xFormers Attention (\texttt{xformers.ops.memory\_efficient\_attention})  
\end{itemize}
By intercepting $QK^\top$ at these kernels, attention maps can be extracted even in models lacking explicit \texttt{attn2}.  
Nevertheless, in DiT (token-concatenation conditioning), the resulting activations remain flat.

\end{document}